# PECULIARITIES OF THERMODYNAMIC SIMULATION WITH THE METHOD OF BOUND AFFINITY.

B. Zilbergleyt,
System Dynamics Research Foundation, Chicago,
livent@ameritech.net

ABSTRACT.
Thermodynamic simulation of chemical and metallurgical systems is the only method to predict their equilibrium composition and is the most important application of chemical thermodynamics. Regardless the basic approach – Gibbs' free energy minimization, maximization of the reaction entropy, or solution to the equations for equilibrium constants – the conventional strategy is always to find the most probable chemical and phase composition of the system, corresponding to the so called true thermodynamic equilibrium. Traditional simulation methods do not account for interactions within the chemical system. The Method of Bound Affinity (MBA[i]) is based on the theory that explicitly takes into account interactions between subsystems of a complex chemical system and leads sometimes to essential differences in simulation results. This article discusses peculiarities of MBA application, exemplified by results for a complex system with a set of subsystems.

INTRODUCTION.

Practical application of chemical thermodynamics is related to a great extent to thermodynamic simulation of the chemical systems. The goal of the simulation is to find the most probable chemical and phase composition of the equilibrium system. Maximum probability corresponds to thermodynamic equilibrium, the only state recognized as equilibrium by current theory. The well-known Zeldovitch's theorem [1], asserting a uniqueness of the equilibrium state for isolated system, forms a cornerstone of the application, thus prompting us to search that unique point in the space of states of the chemical system. In the most of the simulation software developed up to now [e.g., 2,3,4], the standard search usually consists of finding the global minimum of the system's Gibbs' free energy.

Recently introduced ideas in the thermodynamics of chemical equilibrium, presenting the equilibrium as a balance of the thermodynamic forces [5,6], explicitly account for the interactions between subsystems of a complex system, leading to more detailed and accurate evaluation of the system's composition in chemical equilibrium. At constant pressure and temperature, chemical equilibrium follows the logistic equation

$$\Delta G_j^*/RT - \tau_j \delta_j^* (1-\delta_j^*) = 0. \qquad (1)$$

Here $\delta_j^*$ is a subsystem's shift from thermodynamic equilibrium in terms of reaction extent, and *reduced chaotic temperature* $\tau_j$ is the ratio between intensity of the external impact on the system, expressed by the value of alternative temperature [6], and its ability to resist to any shifts from equilibrium due to that interaction, expessed by thermodynamic temperature. An appropriate diagram, constructed from solutions of equation (1), forms the domain of states of the chemical system, and has 4 consecutive typical characteristic areas. For thermodynamic simulation, two leftmost areas (see. Fig.1) are of the most practical interest – the area of thermodynamic equilibrium (TdE) housing the states with zero deviation from thermodynamic equilibrium, and the area of the open equilibrium (OpE) where the subsystem's state deviates from TdE but still dwells on the thermodynamic branch. The parabolic term of equation (1) equals to zero within the TdE area due to $\delta_j^*=0$; equation (1) turns into traditional equation for equilibrium constant, the system may be considered isolated, and traditional thermodynamic simulation is applicable there. The second is the area of open equilibrium where $\delta_j^* \neq 0$. The system ceases to be isolated; now the parabolic term should be taken into account. Beyond the OpE area, the thermodynamic branch becomes unstable, and there follow areas of bifurcations and then chaos. The name of parameter $\tau_j$ reminds that its increase leads the chemical system through those areas. The area limits in term of $\tau_j$ depend upon the "strength", i.e. the value and the sign of standard change of Gibbs' free energy of the chemical reaction, running within the system.

---

[i] In some previous publications we called it a Method of Chemical Dynamics. Currently it seems more reasonable to return to the original name of the method, the Method of Bound Affinity (MBA) [7].

SOME FEATURES OF THE MBA.

In following discussion we will use equation (1) in the reduced by RT form (2), that is as a logistic equation with negative feedback

$$\ln[\Pi_j(\eta_j,0)/\Pi_j(\eta_j,\delta_j^*)] - \tau_j \delta_j^*(1-\delta_j^*) = 0. \qquad (2)$$

The system's deviation $\delta_j^*$ from thermodynamic equilibrium is positive if the system state shifted towards initial reacting mixture, and negative for the opposite direction; $\Pi_j(\eta_j,\delta_j^*)$ is the mole fractions product of the j-reaction participants. The asterisk relates the values to the state of chemical equilibrium. Another possible option, not discussed in this article, is minimization of the systems' Gibbs' free energy, which means

$$G_j^* = \Sigma(n_{kj}^*)\mu_k^0/RT + \Sigma(n_{kj}^*)\ln\Pi_j(\eta_j,\delta_j^*) + \tau_j(\delta_j^*)^2/2 = \min. \qquad (3)$$

It is easy to see that if $\delta_j^*=0$, equations (1), (2) and (3) correspondingly turn into the classical equations for the standard change and absolute value of system's Gibbs' free energy. A set of typical domains of states for a simple reaction

$$PCl_3 + Cl_2 = PCl_5 \qquad (4)$$

is shown in Fig. 1. The fractions shown at the curves indicate the thermodynamic equivalent values of transformation $\eta_j$ [6] at different temperatures, that is, the mole amounts per stoichiometric unit transformed on the reaction path to equilibrium.

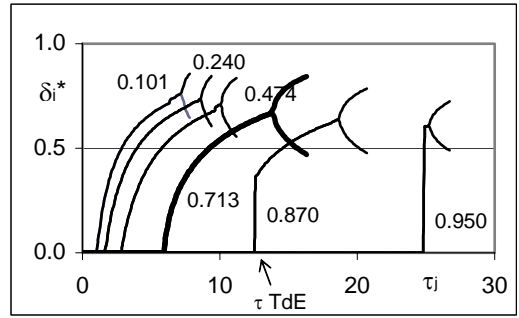

Fig. 1. The part of the domain of states with positive deviations from TdE, reaction (4).

The curves from zero point of the reference frame up to the fork bifurcations corresponds to the thermodynamic branch of evolution of the chemical system [8], and is obviously broken by 2 parts - the horizontal and ascending. The horizontal part is TdE area, and the ascending parts before bifurcations are related to the *open equilibria* (OpE). At the first bifurcation point, which is the OpE limit, the thermodynamic branch decays, splitting into two new non-stable branches. Further increase of the $\tau_j$ value leads to doubling of bifurcation period and eventually to chaos. Both equilibrium areas are typical for any direction of the system's deviation from equilibrium, while bifurcations were found only in cases of shifts toward initial reacting mixtures ($\delta_j^*=0$) as indicated in Fig. 2.

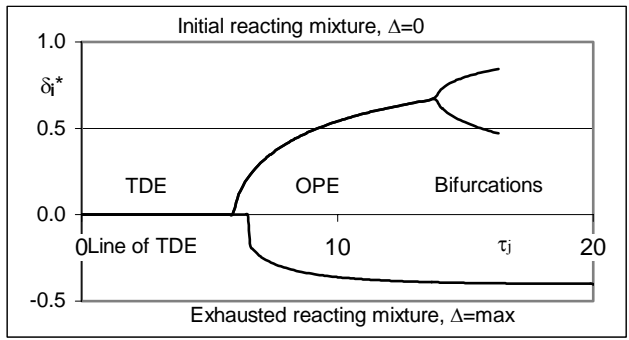

Fig. 2. Two-way bifurcation curve for $\eta_j=0.713$, reaction (4).

It is obvious from equations (1) and (2) that, depending upon the system states location in the domain, we have either classical case of isolated equilibrium with $\delta_j^*=0$ where one doesn't have to take into account $\tau_j$ in the calculations, or open equilibrium where we have to account for $\tau_j$. The TdE limit defines the transition from the classical, conventional simulation to the non-classical approach. Regardless of the simulation problem, it is very important to know the domain of states characteristic points $\tau_{TdEj}$ and $\tau_{OpEj}$, in order to apply the proper algorithms.

Thermodynamic simulation, based on the MBA, includes a search for the area limits and for dependence of $\tau_j$ on deviation from thermodynamic equilibrium within the OpE area. This search includes every subsystem of a complex chemical system at the given initial amount of moles for the participants of j-reaction $n^0_{kj}$, and p, T and $\Delta G^0_j$. When those parameters are found, the thermodynamic simulation consists of the joint solution for the set of logistic equations (2) using corresponding $\tau_j$ values. Another option is to minimize Gibbs' free energy of the subsystems, while also accounting for the subsystem interactions.

AREA LIMITS AND REDUCED CHAOTIC TEMPERATURE.

The area limits may be spotted by direct computer simulation given the initial composition and thermodynamic parameters. At the same time the limits, $\tau_{TdE}$ and $\tau_{OpE}$ can be found analytically with a good precision, avoiding any simulation. Recall that equation (2) contains 2 functions, logarithmic and parabolic. Both have at least one joint point at $\delta_j^*=0$ (Fig. 3) in the beginning of the reference frame, providing for a trivial solution to equation (2) and retaining the system within the TDE area. The curves may cross somewhere else at least one time more; in this case the solution will differ from zero, and number of the roots will be more than one. There is no intersection if

$$d(\tau\delta\Delta)/d\delta < d[\ln(\Pi`/\Pi^*)]/d\delta. \quad (4)$$

This condition leads to a universal formula to calculate the TDE limit as

$$\tau_{TDE}=1+\eta_j\Sigma\,[\nu_{kj}/(n^0_{kj}-\nu_{kj}\eta_j)]\,, \quad (5)$$

where $n^0_{kj}$ is the initial amount and $\nu_{kj}$ is the stoichiometric coefficient of the k-participant in the j-system. We offer this derivation to the reader to check for correctness. Though the area with $\delta_j^*<0$ is more complicated, formula (5) is still valid in cases when the system gets exhausted from one or more of the reactants before the minimum of the logarithmic term occurs. For the reaction A+B=C (exact stoichiometric image of reaction (4)) with initial amounts of participants of respectively 1, 1, and 0 moles, formula (5) may be written down as

$$\tau_{TDE} = (1+\eta_j)/(1-\eta_j). \tag{6}$$

Fig. 4 shows the comparison between the $\tau_{TDE}$ versus $\eta_j$ curve obtained by an iterative process and the curve obtained from calculations using the formulae (5) and (6) for reaction (4). With reference to the OpE limit, it physically means the end of the thermodynamic branch stability where the Liapunov exponent changes its value from negative to positive, and the iterations start to diverge. If the logistic equation (2) is written in the form of

$$\delta_{j(n+1)}^{*} = f(\delta_{jn}^{*}), \tag{7}$$

the OpE limit can be found as a point along the $\tau_j$ axis where the $|f`(\delta_{jn}^{*})|$ value changes from (-1) to (+1) [9]. As of now, we do not have ready formula for this limit.

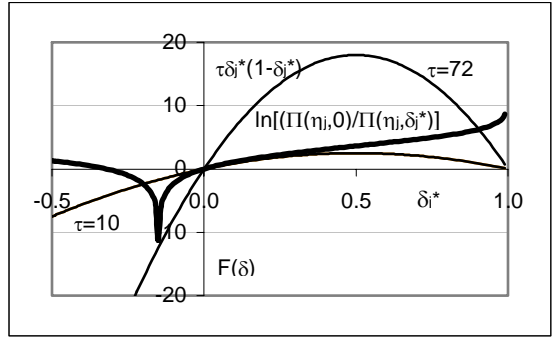

Fig. 3. Terms of equation (1) calculated for reaction (3), $\eta_j$=0.87 (T=348.15K). Logarithmic term curve in bold.

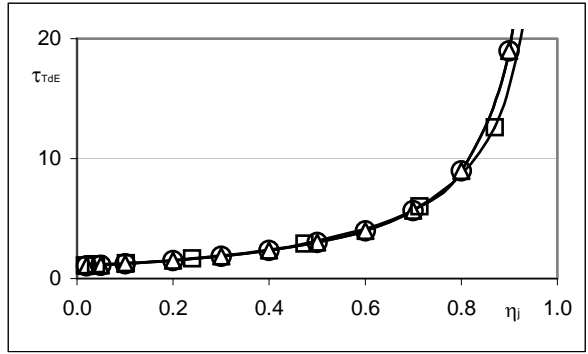

Fig. 4. Calculated and simulated values $\tau_{TDE}$ vs. $\eta_j$. Series o, $\Delta$ and $\square$ represent results for reaction (3) calculated by equation (5), equation (6) and simulated, correspondingly.

The dependence of the area limits upon the standard change of Gibbs' free energy for reaction (4) is shown in Fig. 5. Apparently, the "stronger" the reaction in the system, i.e. the more negative its standard change of Gibbs' free energy, the more extended is its TdE area along the $\tau_j$ axis; both limits, TdE and OpE get closer with the increase of negative $\Delta G^0_j$, merging at its extreme for this reaction value.

To perform calculations in the OpE area one has to know the value of $\tau_j$. The phenomenological theory offers several ways to find it. In the more general method, $\tau_j$ can be found directly from equation (1) as

$$\tau_j = \ln[\Pi_j(\eta_j, 0)/\Pi_j(\eta_j, \delta_j^{*})] / [\delta_j^{*}(1-\delta_j^{*})]. \tag{8}$$

Varying the $\eta_j$ and $\delta_j^{*}$ values, one can create the domain of states as a reference table allowing

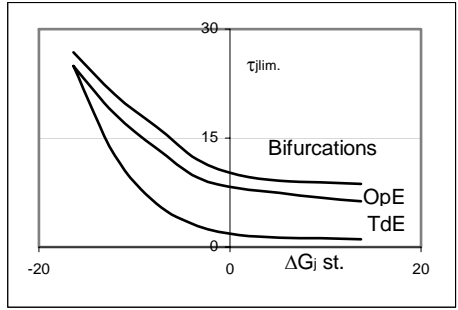

Fig.5. The area limits on the bifurcation diagram $\tau_{jlim.}$ vs. $\Delta G^0_j$, kJ/m, reaction (3).

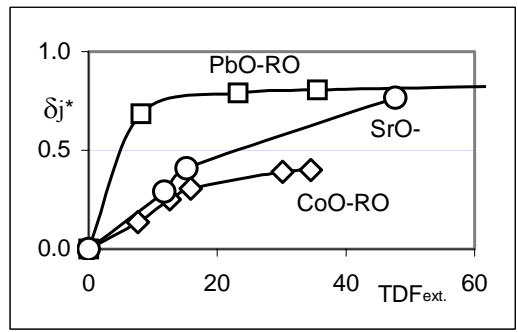

Fig. 6. Shift vs. TDF in homological series of double oxides, reaction of the double oxides with sulfur, HSC simulation.

for extraction of $\tau_j$ during simulation, depending upon the first two parameters.

An alternative method consists of finding the equilibrium composition and the appropriate $\eta_j$ and $\delta_j^*$ values in the homological series of chemical reactions by varying the external TDF [5]. This point-by-point method is illustrated by Fig. 6, where ▪RO stands for restricting oxide of the couple. The standard change of Gibbs' free energy of the double oxide formation from oxides was accepted as the TDF.

Needless to say, prior to the search for $\tau_j$ one has to find $\eta_j$ for the reaction in question at given temperature. It can be done by any traditional simulation method for thermodynamic equilibrium (at $\delta_j^*=0$).

EXAMPLES.

The MBA obviously consists of 2 steps – mandatory solutions for isolated subsystems precede solutions for the interacting subsystems.

The following example is a graphical image of domains of states for a hypothetical chemical system with 5 subsystems. We tried to show possible different locations of the subsystems in their domains of states in equilibrium. The internal pentagon connects the TdE limit values of the subsystems, the external pentagon does the same for OpE limits; the pentagon with the marks connects conditionally simulated $\tau_j$ values.

How different may be results of conventional and MBA thermodynamic simulation? To answer this question, we used several chemical systems with 2 subsystems where different reactions between a couple of double oxides and hydrogen can run. Double oxides were taken to reduce reaction activity of the basic oxides, in other words, to make the reaction less "strong", allowing usage of the non-zero parabolic term in the OpE area in calculations. For example, binding of CoO into CoO▪TiO2 lowers equilibrium reaction constant from 32.09 for

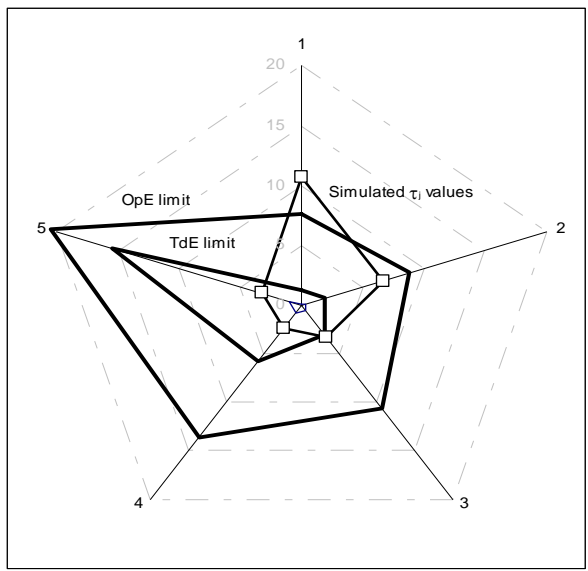

Fig. 7. A flat radar presentation of the sub-system domains of states for a hypothetic chemical system. Subsystem numbers are shown at the beams. (This image is flat!)

$CoO+H_2=Co+H_2O$ down to 0.684 for $CoO \cdot TiO_2+H_2=Co+TiO_2+H_2O$. Simulation was carried using HSC and then MBA in mixtures of double oxides ($Me_1O \cdot TiO_2+Me_2O \cdot TiO_2$) with hydrogen at T = 973.16 and p = 0.1MPa; results are shown in Table 1. We used the following values of $\tau_j$, found by the point method for MeO and MeO·TiO2: Reactions NiO and NiO·TiO2 with hydrogen $\tau_j$ = 29.04; for CdO-reactions $\tau_j$ = 21.13; for CoO-reactions $\tau_j$ = 18.57.
The difference between results of classical and MBA simulations is quite discernable.

Table 1.
Comparison between results, obtained by conventional (HSC) and MBA simulations.

| Calculated equilibrium contents, moles. | | | | | | | | |
|---|---|---|---|---|---|---|---|---|
| *Couple* | HSC | MBA | *Couple* | HSC | MBA | *Couple* | HSC | MBA |
| NiO·TiO2 | 0.126 | 0.164 | NiO·TiO2 | 0.088 | 0.157 | CdO·TiO2 | 0.208 | 0.258 |
| CdO·TiO2 | 0.322 | 0.276 | CoO·TiO2 | 0.488 | 0.389 | CoO·TiO2 | 0.437 | 0.380 |

CONCLUSION.

Described peculiarities of the Method of Bound Affinity at constant temperature and pressure and ways to find area limits and running values of $\tau_j$, allow for application of the MBA to thermodynamic simulation of chemical systems. Results by a classical simulation method and the MBA are quite different; the new theory prompts us to consider the MBA results to be in better correspondence with real chemical systems. Some results of this work were presented in part earlier on [10].

REFERENCES.